**Effect of nanoclay laponite and pH on the energy transfer between fluorescent dyes**

Dibyendu Dey, D. Bhattacharjee, S. Chakraborty, Syed Arshad Hussain*

Department of Physics, Tripura University, Suryamaninagar – 799022, Tripura, India

* Corresponding author

Email: sa_h153@hotmail.com, sah.phy@tripurauniv.in

Ph: +919862804849 (M), +91381 2375317 (O)

Fax: +913812374802

**Abstract:**

Fluorescence resonance energy transfer (FRET) between two dyes acriflavine (Acf) and rhodamine B (RhB) were investigated in solution and Layer-by-Layer (LbL) self assembled films in presence and absence of clay mineral laponite. UV-Vis absorption and fluorescence spectroscopy studies suggest both the dyes present mainly as monomer in solution and films. Energy transfer occurred from Acf to RhB in solution and LbL films. The energy transfer efficiency increases in presence of clay laponite and the maximum efficiency were 78.17% and 32.54% in clay dispersion and in LbL films respectively. Presence of laponite particles onto LbL film was confirmed by atomic force microscopy investigations with a surface coverage of more than 75%. Energy transfer efficiency was pH sensitive and the energy transfer efficiency varies from 4.5% to 44.45% in mixed dye solution for a change in pH from 3.0 to 12.0. With proper calibration it is possible to use the present system under investigation to sense pH over a wide range of pH from 3.0 to 12.0.

# 1. Introduction

Fluorescence resonance energy transfer (FRET) is an electrodynamic phenomenon that can occur through the transfer of excited state energy from donor to acceptor. The theoretical analysis was well developed by Förster [1, 2]. The rate of energy transfer depends upon the extent of spectral overlapping area of the fluorescence spectrum of donor with the absorption spectrum of the acceptor, the relative orientation of the donor and acceptor transition dipoles and the distance between these molecules [1-3]. Due to its sensitivity to distance, FRET has been used to investigate molecular level interaction [3-10]. Fluorescence emission rate of energy transfer has wide applications in biomedical, protein folding, RNA/DNA identification and their energy transfer process [4-10].

FRET mechanisms are also important to other phenomena, such as photosynthesis kinetics, chemical reactions and Brownian dynamics [11, 12]. Another important application of FRET phenomenon is in dye lasers. If a dye laser has to be used as an ideal source, its spectral region needs to be extended. The use of such energy transfer in dye lasers is also helpful in minimizing the photo quenching effects and thereby increasing the laser efficiency.

On the other hand clay platelets are natural nano-particles with layered structure. Due to the cation exchange capacity (CEC) of clay, the dye molecules are adsorbed onto the clay surfaces [13, 14]. Dye adsorption enhances the concentration of the dye molecules, which may promote their intermolecular physical and chemical interactions. For example, if two molecules are in close proximity, fluorescence resonance energy transfer may occur. Probably first record on efficient energy transfer in clay mineral systems, based on the interaction between two different dyes are clay mineral dispersions with cyanine and rhodamine dyes simultaneously adsorbed on clay mineral surfaces [14, 15]. Further examples of energy transfer in clay mineral systems are triplet-triplet energy transfer from bound sensitizers to mircene [16] and to aromatic hydrocarbons adsorbed in hydrophobic organo-clay [17]. It was observed that the clay/porphyrin complexes are promising and prospective candidates to be used for construction of efficient artificial light-harvesting system [18]. Czímerová et al.[19] reported prominent energy transfer among laser dyes in saponite dispersion. The FRET between cationic polypeptide polylysine and cyanine dyes was reported in LbL films of clay minerals [20]. J. Bujdák et al studied the FRET between two Rhodamines Rh123 (donor) and Rh610 (acceptor) in both solution and in presence of nanoclay saponite (SAP) [21]. It was observed that the FRET efficiency was higher in presence of SAP. The clay mineral works as templates for concentrating the dyes, accordingly the intermolecular separation between them decreases. To avoid the aggregation and the fluorescence self quenching of the dyes, a hydrophobic surfactant was introduced which suppressed the aggregation of the dyes [21]. In another work by the same group, FRET phenomenon between laser dyes rhodamine 123 (R123), rhodamine 610 (R610), and oxazine 4 (Ox4) has been reported. The dye molecules played the role of molecular antennas and energy donors (R123), energy acceptors (Ox4), or both (R610). It was observed that the FRET efficiency increases in presence of laponite [22]. In one of our previous work we have observed that the energy transfer efficiency among dyes assembled in Langmuir-Blodgett films increases in presence of clay sheets [3].

In the present communication, the FRET phenomenon between two laser dyes acriflavine (Acf) and rhodamine B (RhB) has been reported. We investigated this phenomenon in aqueous solution, clay dispersion and in Layer-by-Layer (LbL) self assembled films. Also the effects of pH on the FRET efficiency have been investigated. These two dyes Acf and RhB are in principle suitable for FRET [23]. Both the dyes are highly fluorescent. The fluorescence spectrum of Acf sufficiently overlaps with the absorption spectrum of RhB. P. D. Sahare et al [23] observed energy transfer in binary solution mixture of acriflavine and rhodamine 6G and acriflavine and rhodamine B by life time measurement. The spectra of Acf are highly pH sensitive due to the presence of electron donor type functional group [24]. Excitation energy transfer between acriflavine and rhodamine 6G as pH sensor has already been demonstrated [24]. However, effect of nano-clay platelet laponite as well as pH on energy transfer using Acf and RhB has never been studied. Therefore it is very interesting to study the FRET parameters at different pH using Acf as donor in order to explore their possible application as pH sensor. The aim of this study was to investigate the effect of nano- clay platelet laponite and pH on FRET efficiency between these two dyes in solution and in LbL films in order to explore their possible applications.

## 2. Experimental

*2.1. Materials*

Both the dyes Acf and RhB were purchased from Sigma Chemical Co., USA and used as received. Molecular structures of the dyes are shown in the inset of figure 1. Millipore water or HPLC grade methanol [Acros Organics, USA] were used as solvent. The clay mineral used in the present work was Laponite, obtained from Laponite Inorganic, UK and used as received. The size of the clay platelet is less than 0.05 μm and CEC is 0.739 meq/g determined with CsCl [25]. The clay dispersion was prepared in Millipore water and stirred for 24 h with a magnetic stirrer followed by 30 minute ultrasonication before use. Poly (acrylic acid) (PAA) and poly (allylamine hydrochloride) (PAH) were used as polyanion and polycation during Layer-by-Layer (LbL) self assembled film preparation. Both PAA and PAH were purchased from Aldrich Chemical Co., USA and was used without further purification.

*2.2. Film preparation*

Electrolyte deposition bath of cationic dye RhB and Acf were prepared with $10^{-4}$M aqueous solution using triple distilled deionized (resistivity 18.2 MΩ-cm) Millipore water. The anionic electrolytic bath of PAA was prepared also with triple distilled deionized Millipore water (0.25 mg/ml). LbL self assembled films were obtained by dipping thoroughly clean fluorescence grade quartz substrate alternately in solutions of anionic PAA and oppositely charged RhB and Acf dye mixtures. LbL method utilizes the Vander Waals interactions between the quartz slide and PAA as well as charge transfer interaction between PAA and cationic dyes [26, 27]. The quartz slide was dipped in the aqueous solution of PAA for 30 minutes. Then it was taken out and sufficient time was allowed for drying and then rinsing in water bath for 2 minutes so that the surplus anion attached to the surface washed off. The dried substrate was then immersed in cationic dye mixture (RhB + Acf) followed by same rinsing procedure. Deposition of PAA and RhB and Acf layers resulted in one bi-layer of self assemble film. The incorporation of clay in the LbL film was done with the help of aqueous PAH solution (0.25 mg/ml). For this the quartz slide was first dipped in electrolytic polycation (PAH) aqueous solution for 30 minutes followed by same rinsing in water bath and drying procedure and then dipped into the anionic clay dispersion which is again followed by rinsing action in water bath. The slide thus prepared was dipped in the cationic electrolytic solution of RhB and Acf. Due to electrostatic interaction cationic Acf and RhB were adsorbed onto the negative charged surface of the clay in LbL films.

*2.3. UV–Vis absorption and fluorescence spectra measurement*

UV–Vis absorption and steady state fluorescence spectra were recorded by a Perkin Elmer UV-Vis Spectrophotometer (Lambda-25) and a Perkin Elmer Fluorescence

Spectrophotometer (LS-55) respectively. The fluorescence light was collected from the sample surface at an angle of 45° (front face geometry) and the excitation wavelength was 420 nm.

*2.4. Theoretical considerations*

Solving the enigma surrounding fluorescence quenching experiments revealed the phenomenon of FRET and led J. Perrin [28] to propose dipole–dipole interactions as the mechanism via which molecules can interact without collisions at distances greater than their molecular diameters. Some 20 years later, Förster [1, 2] built upon Perrin's idea to put forward an elegant theory which provided a quantitative explanation for the non-radiative energy transfer in terms of his famous expression given by

$$k_T(r) = \frac{1}{\tau_d}\left(\frac{R_0}{r}\right)^6$$

Where, $k_T(r)$ is the rate of energy transfer from donor to acceptor and $R_0$ is the well-known Förster radius given by the spectral overlap between the fluorescence spectrum of the donor and the absorption spectrum of the acceptor. Since then the technique of FRET has come a long way finding applications in most of the disciplines, which by itself signifies the importance of Förster's formulation and usefulness of this technique.

In the present manuscript based on Förster's theory the spectroscopic results were used in order to quantify the FRET parameters between two dyes under investigations. The details of the background theory for calculating different FRET parameters have been given in the supporting information.

**3. Results and discussion**

*3.1. The UV-Vis absorption and steady state fluorescence spectroscopy*

Normalized UV-Vis absorption and steady state fluorescence spectra of pure Acf and RhB in aqueous solutions are shown in figure 1. Both absorption and fluorescence spectra are characteristics of the presence of monomers. The fluorescence spectra were recorded by exciting the corresponding absorption maxima of Acf and RhB. The absorption and fluorescence maxima of Acf are centered at 449 and 502 nm respectively which is assigned due to the Acf monomers [23]. Acf monomer absorption band within 444 – 453 nm depending on the concentration has been reported [29]. For Acf dimer it has been reported that instead of a single monomer band two bands at around 437 and 470 nm are observed with the intensity of the blue band higher than the other [29].

On the other hand RhB absorption spectrum possess prominent intense 0-0 band at 553 nm along with a weak hump at 520 nm which is assigned due to the 0-1 vibronic transition [30]. Similar reports with RhB monomer bands at 553 nm and 0-1 vibronic components of monomer at 525 nm have been reported. [31]. For J dimer of RhB the absorption bands are found to be red shifted to 569 and 531 nm [31] However, for H-dimer the dominance of 531 nm band intensity with respect to the intensity of 553 nm band have been reported [31, 32]. The RhB fluorescence spectrum shows prominent band at 571 nm which is assigned due to the RhB monomeric emission [30].

A close look at the figure 1 reveals that there exists sufficient overlapping of Acf fluorescence spectrum and RhB absorption spectrum. This justifies the selection of these two dyes in order to quantify energy transfer from Acf to RhB. Here Acf acts as a donor and RhB acts as an acceptor. Also both the dyes are highly fluorescent, which are the prerequisite for FRET to occur [1-3].

*3.2. FRET between Acf and RhB*

*3.2.1. Solution and clay dispersion*

In order to investigate the possible FRET between Acf and RhB, the fluorescence spectra of Acf, RhB and their mixture in different conditions are measured with exciting wavelength at 420 nm. The excitation (absorption) wavelength was selected approximately to excite the Acf molecules directly and to avoid or minimize the direct excitation of the RhB molecules.

Figure 2a shows the fluorescence spectra of pure Acf, RhB and their mixture (50:50 volume ratios) in aqueous solution as well as in clay dispersions. Figure 2a reveals strong prominent Acf fluorescence band where as the RhB fluorescence band is very less in intensity in case of pure dye solution. The less intensity of pure RhB fluorescence band indicates very small contribution of direct excitation of the RhB molecules. The fluorescence spectra of Acf-RhB mixture is very interesting, here the Acf fluorescence intensity decreases in favour of RhB fluorescence band. This decrease in Acf fluorescence intensity is due to the transfer of energy from Acf to RhB molecules. This transferred energy excites more RhB molecules followed by light emission from RhB, which is added to the original RhB fluorescence. As a result the RhB fluorescence intensity gets sensitized. Inset of figure 2a shows the excitation spectra measured with excitation wavelength fixed at Acf (500 nm) and RhB (571 nm) fluorescence maxima in case of Acf-RhB mixed aqueous solution. Interestingly both the excitation spectra are almost similar and possess characteristic absorption bands of Acf monomers. This confirms that the RhB fluorescence is mainly due to the light absorption by Acf and corresponding transfer to RhB monomer. Thus FRET between Acf to RhB has been confirmed.

In order to check the effect of nanoclay platelets on FRET, we measured the fluorescence spectrum of pure Acf, RhB as well as Acf-RhB mixture in laponite clay dispersion [figure 2a].

Intensities and band positions for Acf and RhB fluorescence are listed in table 1. A red shift of RhB fluorescence band of the order of 8 nm occurred. Such smaller shift in RhB fluorescence in montmorillonite and hectorite were reported and attributed to monomer fluorescence of RhB adsorbed on the external clay surface or in the interlamellar regions of clay sheets [14]. In the present case we also consider that clay dispersions mainly contain the RhB monomer and the observed shift is due to the consequent adsorption of dye molecules on the clay surface and in the interlamelar space of the clay sheets. Also the intensity for both Acf and RhB fluorescence decreases in clay dispersion. This decrease in fluorescence intensity is attributed to the scattering of light from the clay particles present in the dispersion [19]. The most interesting observation in clay dispersion was that the Acf fluorescence intensity decreases further in favour of RhB fluorescence intensity in presence of nanoclay platelets (figure 2a, curve 6), results an increase in FRET efficiency.

It is worthwhile to mention in this context that clay particles are negatively charged and have layered structure with a cation exchange capacity [14, 15]. Both the dyes Acf and RhB under investigation are positively charged. Accordingly they are adsorbed onto the clay layers [14, 15]. On the other hand FRET process is very sensitive to distances between the energy donor and acceptor and occurs only when the distance is of the order of 1-10 nm [1-3]. Therefore, in the present case, clay particles play an important role in determining the concentration of the dyes on their surfaces or to make possible close interaction between energy donor and acceptor in contrast to the aqueous solution. In one of our previous work we demonstrated the enhancement of FRET efficiency between two dyes in presence of nanoclay sheet laponite [3].

Analysis of fluorescence spectra (figure 2a) reveal that the spectral overlapping integral $J(\lambda)$ between the fluorescence spectrum of Acf (donor) and absorption spectra of RhB (acceptor) as well as energy transfer efficiency increases due to incorporation of nanoclay sheets (table 2). Also due to the presence of nanoclay sheet laponite, the intermolecular distance between Acf and RhB decreases from 8.07 nm to 5.33 nm. So clay particles play a vital role in concentrating the dyes on their surfaces and thus reducing the intermolecular distance providing a favourable condition for efficient energy transfer. Consequently the energy transfer efficiency increases from 11.37% to 78.17% in presence of clay platelets.

In order to check the effect of donor / acceptor concentration on FRET, fluorescence spectra of Acf-RhB mixture in presence of clay platelet laponite were measured with fixed amount of Acf (donor) and varying amount of RhB (acceptor). Figure 2b shows the fluorescence spectra of Acf-RhB mixed dye system with fixed amount of Acf and varying amount of RhB in presence of clay platelet laponite. The values of spectral overlap integral ($J(\lambda)$), energy transfer efficiency (E), Förster radius ($R_0$) and the donor acceptor distance (r) calculated from figure 2b and listed in table 3. Interestingly it was observed that for a fixed amount of donor Acf the FRET efficiency increases with the increase in acceptor concentration in the Acf-RhB mixture. Maximum energy transfer efficiency (table 3) was 78.17% for acceptor concentration of 50%.

*3.2.2. Layer-by-Layer self assembled films*

Figure 3a shows the fluorescence spectra of pure RhB, Acf and their mixture (50:50 volume ratios) in 1 bi-layer LbL self assembled films in presence and absence of clay particles. Here also pure Acf shows strong fluorescence with monomer band at 523 nm, both in presence and absence of clay, which is red shifted with respect to aqueous solution or clay dispersion. This shift may be due to the change in microenvironment when Acf molecules are incorporated into the polymer (PAH and PAA) backbone in the restricted geometry of solid surface during LbL film formation. For RhB the trend is very similar to clay dispersion and shows very weak fluorescence with peak at around 575 nm. Energy transfer is observed for mixed dye system in LbL film. However, the energy transfer efficiency is less compared to their solution counterpart. This observed decrease in energy transfer efficiency with respect to solution may be due to the observed small value of the overlap integral ($J(\lambda)$) in LbL films. However, in presence of clay the efficiency increases in LbL films.

Fluorescence spectra of AcF-RhB mixed LbL films prepared with fixed amount of Acf and varying amount of RhB in presence of clay is shown in figure 3b. The corresponding energy transfer efficiency (E), Förster radius ($R_0$), spectral overlapping integral ($J(\lambda)$) and the distance between the donor and acceptor (r) for Acf-RhB mixed LbL films are listed in table 4. Here the trend is very similar to that of clay dispersion. Here also the energy transfer efficiency increases with increase in acceptor concentration in the mixed films. The maximum FRET efficiency was 32.54% for an acceptor concentration of 50% in the mixed LbL films.

*3.3. Effect of pH on FRET*

Among the molecules under current investigation Acf is pH sensitive because of its basic nature of the central nitrogen atom [33]. The fluorescence spectra of Acf are affected with change in pH [24]. This may in turn cause a change in spectral overlapping of the donor fluorescence and acceptor absorbance resulting a change in FRET efficiency. In order to check the effect of pH on FRET process, fluorescence spectra of Acf-RhB mixture in aqueous solution prepared at different pH were measured (figure available in supporting information). It was observed that the Acf fluorescence was red shifted with decrease in pH.

It is interesting to mention in this context that proflavine molecule is very similar to acriflavine with regards to protonation and deprotonation. Proflavine has been found to exist as single protonated, double protonated as well as neutral molecules in aqueous solution with pKa ~ 9.5 for single protonated and 0.2 for double protonated form [34, 35]. The excited state dissociation constants are 12.5 for single protonated and 1.5 for double protonated species. It has been observed that acriflavine mainly remain as double protonated form in Nafion (a perfluorosulfonate cation exchange membrane) due to the high local proton concentration.[35] Larger red shift in Acf fluorescence in nafion has been observed and explained due to change in the dipole moments in the excited state of the double protonated Acf [34] and due to the broad distribution of pKa in nafion matrix [34]. In the present case at lower pH red shift of Acf fluorescence is observed. At lower pH Acf molecules mainly remain as double protonated form due to the increase in local proton concentration with decreasing pH. Accordingly the dipole moments of the excited state of double protonated Acf have been changed. This change in dipole moments may be responsible for the observed large stoke shift / red shift of the Acf fluorescence.

The values of $J(\lambda)$ and FRET efficiencies calculated from the spectra measured at different pH are listed in table 5. Interestingly it was observed that the FRET efficiency increases with increase in pH. It was found that for the same donor acceptor concentration and excitation wavelength, the value of spectral overlap integral $J(\lambda)$ changes a lot with change in pH. But the shape of the fluorescence spectra remains almost similar.

It is worthwhile to mention in this context that RhB contains – COOH group, which can dissociate in certain conditions to form cations – anions (zwitterions). In basic medium RhB shows the zwitterionic form which could be responsible for the close interaction between cationic acriflavine ($Acf^+$) and $COO^-$ group of zwitterionic RhB. This will increase the possibility of closer approach of Acf and RhB at higher pH resulting an increase in FRET efficiency. In acidic medium (lower pH) RhB generally remains in cationic form with lower pKa value [39]. Also the shift of Acf fluorescence with pH results a change in spectral overlap between Acf fluorescence and RhB absorbance i.e. $J(\lambda)$ value. This will in turn effect the FRET efficiency.

The electron donor type functional group of Acf become more basic with increase in pH in the excited state, consequently the fluorescence spectra shifts towards shorter wavelength providing a larger value of spectral overlap integral (table 5) with increasing pH. This increase in $J(\lambda)$ in turn causes an increase in FRET efficiency. The value of $J(\lambda)$ changes from $18 \times 10^{15}$ $M^{-1}cm^{-1}nm^4$ to $47.7 \times 10^{15}$ $M^{-1}cm^{-1}nm^4$ for change in pH from 3.0 to 12.0. Accordingly the energy transfer efficiency varies from 4.5% to 44.45%. Therefore in the present system under investigation the FRET process between Acf and RhB is very pH sensitive.

Figure 4a and 4b show the plot of spectral overlap integral $J(\lambda)$ and energy transfer efficiency (E) as a function of pH. Interestingly it was observed that both $J(\lambda)$ and E increases almost linearly with increasing pH. Therefore, pH dependence of the energy transfer between the present donor – acceptor pair Acf and RhB under investigation makes the system a suitable candidate for sensing pH. Any of the data from table 5 may be used to sense the pH with appropriate calibration.

It is interesting to mention in this context that energy transfer has already been used for pH measurement [24]. Chan et al. demonstrated Förster resonance energy transfer (FRET)-based ratiometric pH nanoprobes where they used semiconducting polymer dots as a platform. The linear range for pH sensing of the fluorescein-coupled polymer dots was between pH 5.0 and 8.0 [36]. Egami et al. has introduced a fiber optic pH sensor, using polymer doped with either congo red (pH range 3 to 5) or methyl red (pH range from 5 to 7) [37]. pH sensor based on the measurement of absorption of phenol red has also been reported [38], which can sense a pH range of 7 – 7.4. In the present system of pH measurement using the change in FRET parameter with pH is capable of measuring over a wide range of pH 3.0 to 12.0. This is one advantage with respect to previous system.

*3.4. Atomic Force Microscopy*

To confirm the incorporation of clay particles onto LbL films and to have idea about the structure of the film, LbL film was studied by Atomic Force Microscope (AFM). Figure 5 shows a typical AFM image of PAH-laponite-Acf-RhB hybrid LbL film deposited on a Si substrate along with the line analysis spectrum. In the figure, the laponite particles are clearly visible. The hybrid film consists of a close-packed array of hybridized laponite particles. The surface coverage is more than 75%. Few overlapping of laponite particles are also observed. White spots are indicative of aggregates of laponite particles; while some uncovered regions are also observed. Individual laponite particles are not clearly resolved due to the high layer density. From the height profile analysis, it is seen that the height of the monolayer varies between – 2 nm and +2 nm. This includes the height of the PAH layer on substrate plus the height of the laponite layer, and Acf & RhB molecules adsorbed onto the clay surfaces. It is worthwhile to mention in this context that AFM image of Acf - RhB LbL film without clay shows a smooth surface indicating the uniform deposition of dyes without any aggregates [figure not shown]. Since the dimension of the individual dye molecules are beyond the scope of resolution, hence its not possible to distinguish individual Acf or RhB molecules. Therefore, as a whole the AFM investigation give compelling visual evidence of incorporation of laponite particles onto the LbL films.

## 4. Conclusion

Fluorescence resonance energy transfer (FRET) between two fluorescent dyes acriflavine and rhodamine B were investigated successfully in solution and Layer-by-Layer (LbL) self assembled films in presence and absence of clay mineral particle laponite. UV-Vis absorption and fluorescence spectroscopy studies reveal that both the dyes present mainly as monomer in solution and films and there exist sufficient overlap between the fluorescence spectrum of Acf and absorption spectrum of RhB, which is a prerequisite for the FRET to occur from Acf to RhB. Energy transfer occurred from Acf to RhB in both solution and LbL films in presence and absence of laponite. The energy transfer efficiency increases in presence of clay laponite in both solution and in LbL films. The maximum efficiencies were found to be 78.17% and 32.54% for the mixed dye system (50% RhB + 50% Acf) in clay dispersion and LbL films respectively. Atomic force microscopy investigations confirmed the presence of laponite particle in LbL films with a surface coverage of more than 75%. Due to the basic nature of the central nitrogen atom Acf is pH sensitive and it was observed that the overlap between Acf fluorescence and RhB absorption spectrum changes with change in pH. Consequently energy transfer efficiency was found to be pH sensitive and varies from 4.5% to 44.45% in mixed dye solution for a change in pH from 3.0 to 12.0. With proper calibration it's possible to use the present system under investigation to sense pH over a wide range of pH from 3.0 to 12.0.

**Acknowledgements**

The author SAH is grateful to DST, CSIR and DAE for financial support to carry out this research work through DST Fast-Track project Ref. No. SE/FTP/PS-54/2007, CSIR project Ref. 03(1146)/09/EMR-II and DAE Young Scientist Research Award (No.

2009/20/37/8/BRNS/3328). We are grateful to Prof. Robert A. Schoonheydt, K. U. Leuven, Belgium for providing the clay samples.

**Table 1** Fluorescence intensity and band position of Acf & RhB mixtures in aqueous solution and in clay dispersion. Excitation wavelength was 420 nm.

| 1:1 volume ratio of Acf & RhB | Acf fluorescence | | RhB fluorescence | |
|---|---|---|---|---|
| | Band position (nm) | intensity | Band position (nm) | intensity |
| Aqueous solution | 502 | 897 | 571 | 58 |
| Clay dispersion | 502 | 756 | 579 | 49 |

**Table 2** Values of spectral overlap integral ($J(\lambda)$), energy transfer efficiency (E), Förster radius ($R_0$), and donor – acceptor distance (r) calculated from the spectral characteristics of figure 2a.

| | $J(\lambda) \times 10^{15}$ $M^{-1}cm^{-1}nm^4$ | E % | $R_0$ (nm) | r (nm) |
|---|---|---|---|---|
| Aqueous solution | 32.17 | 11.37 | 8.43 | 8.07 |
| Clay dispersion | 53.71 | 78.17 | 6.60 | 5.33 |

**Table 3** Values of spectral overlap integral ($J(\lambda)$), energy transfer efficiency (E), Förster radius ($R_0$) and donor – acceptor distance (r) calculated from the spectral characteristics of figure 2b.

| % of acceptor (RhB) | $J(\lambda) \times 10^{15}$ $M^{-1}cm^{-1}nm^4$ | E % | $R_0$ (nm) | r (nm) |
|---|---|---|---|---|
| 20 | 43.25 | 32.41 | 5.87 | 6.63 |
| 30 | 46.87 | 39.15 | 6.10 | 6.56 |
| 40 | 50.52 | 55.02 | 6.38 | 6.16 |
| 50 | 53.73 | 78.17 | 6.60 | 5.33 |

**Table 4** Values of spectral overlap integral ($J(\lambda)$), energy transfer efficiency (E), Förster radius ($R_0$) and donor – acceptor distance (r), calculated from the spectral characteristics of figure 3b.

| % of acceptor (RhB) | $J(\lambda) \times 10^{15}$ $M^{-1}cm^{-1}nm^4$ | E % | $R_0$ (nm) | r (nm) |
|---|---|---|---|---|
| 20 | 15.33 | 8.82 | 2.78 | 4.10 |
| 30 | 18.87 | 17.50 | 2.95 | 3.81 |
| 40 | 22.25 | 26.10 | 3.10 | 3.68 |
| 50 | 25.17 | 32.54 | 3.23 | 3.64 |

**Table 5** Values of spectral overlap integral ($J(\lambda)$), energy transfer efficiency (E), Förster radius ($R_0$) and donor – acceptor distance (r), calculated from the fluorescence spectra of aqueous solution of Acf-RhB mixture measured at different pH.

| pH of AcF solution | $J(\lambda) \times 10^{15}$ $M^{-1}cm^{-1}nm^4$ | E % |
|---|---|---|
| 3.0 | 18.00 | 04.50 |
| 4.5 | 23.80 | 07.20 |
| 6.0 | 30.10 | 14.11 |
| 7.5 | 34.90 | 21.70 |
| 9.0 | 39.40 | 28.50 |
| 10.5 | 43.20 | 35.20 |
| 12 | 47.70 | 44.45 |

**Figure captions**

**Fig. 1** Normalized UV-Vis absorption and fluorescence spectra of Acf and RhB. The overlap between Acf fluorescence and RhB absorption spectra is shown by shaded region. Inset show molecular structure of (a) RhB and (b) Acf. Dye concentration was $10^{-6}$M.

**Fig. 2** (a) Fluorescence spectra of RhB (1), Acf (2), and Acf+RhB (50:50) mixture (3) in aqueous solution and RhB (4), Acf (5), and Acf+RhB (50:50) mixture (6) in clay dispersion. Inset shows the exitation spectra for Acf+RhB mixture with excitation wavelength at 500 (I) and 571 (II) nm. (b) Fluorescence spectra of Acf and RhB mixture in aqueous laponite dispersion at different concentration for fixed amount of donor (Acf). Inset shows the FRET efficiency as a function of acceptor concentration. All the spectra were measured with excitation wavelength 420 nm. Dye concentration was $10^{-6}$M. For clay dispersion the dye loading was 0.1% of CEC of laponite.

**Fig. 3** (a) Fluorescence spectra of RhB (1), Acf (2), Acf+RhB (50:50) mixture (3) in LbL film without clay and RhB (4), Acf (5), Acf+RhB (50:50) mixture (6) in LbL film with clay. (b) Fluorescence spectra of Acf and RhB mixture for fixed amount of donor (Acf) and varying amount of acceptor in LbL films in presence of clay particle laponite. The inset shows the variation of FRET efficiency as a function of acceptor concentration. All the spectra were measured with excitation wavelength 420 nm. For LbL film preparation dye deposition bath was prepared with dye concentration $10^{-4}$M. Clay concentration for clay deposition bath was 2 ppm.

**Fig. 4** Plot of (a) spectral overlap integral $J(\lambda)$ and (b) energy transfer efficiency (E) as a function of pH.

**Fig. 5** AFM image of Acf-RhB mixed LbL film in presence of clay laponite.

**Table 1** Fluorescence intensity and band position of Acf & RhB mixtures in aqueous solution and in clay dispersion. Excitation wavelength was 420 nm.

**Table 2** Values of spectral overlap integral ($J(\lambda)$), energy transfer efficiency (E), Förster radius ($R_0$), and donor – acceptor distance (r) calculated from the spectral characteristics of figure 2a.

**Table 3** Values of spectral overlap integral ($J(\lambda)$), energy transfer efficiency (E), Förster radius ($R_0$) and donor – acceptor distance (r) calculated from the spectral characteristics of figure 2b.

**Table 4** Values of spectral overlap integral ($J(\lambda)$), energy transfer efficiency (E), Förster radius ($R_0$) and donor – acceptor distance (r), calculated from the spectral characteristics of figure 3b.

**Table 5** Values of spectral overlap integral ($J(\lambda)$), energy transfer efficiency (E) calculated from the fluorescence spectra of aqueous solution of Acf-RhB mixture measured at different pH.

**Figures**

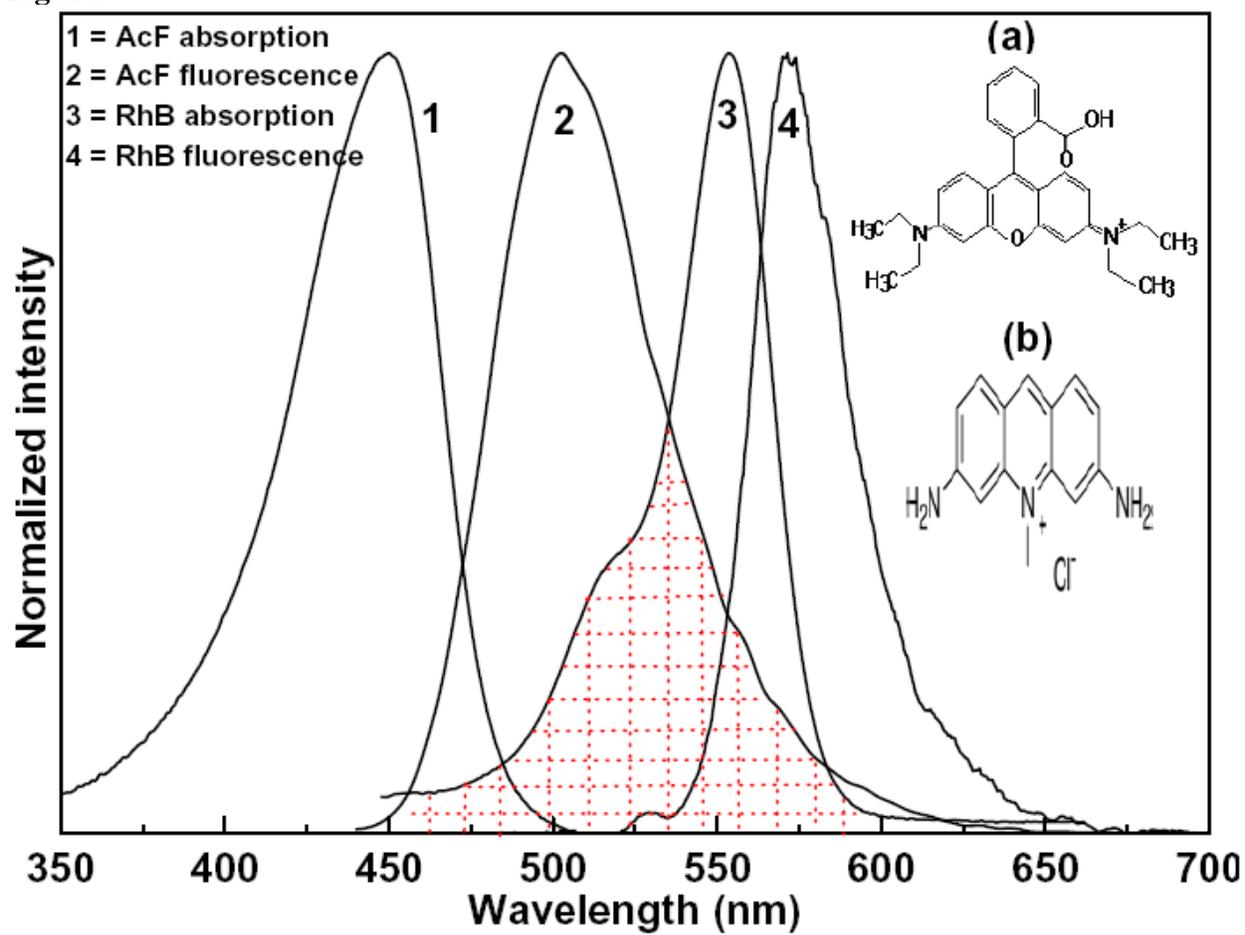

**Fig. 1**

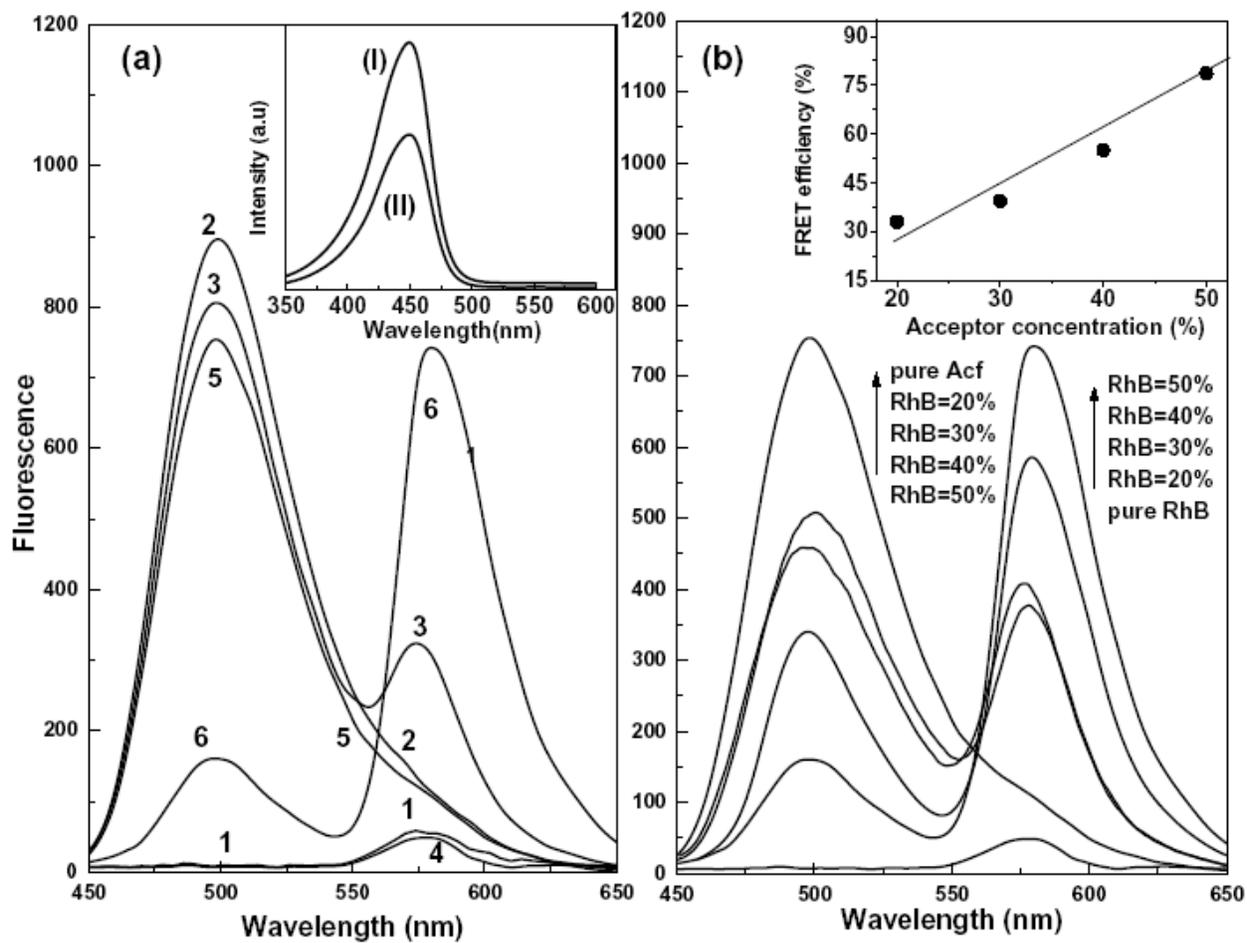

**Fig. 2a and 2b**

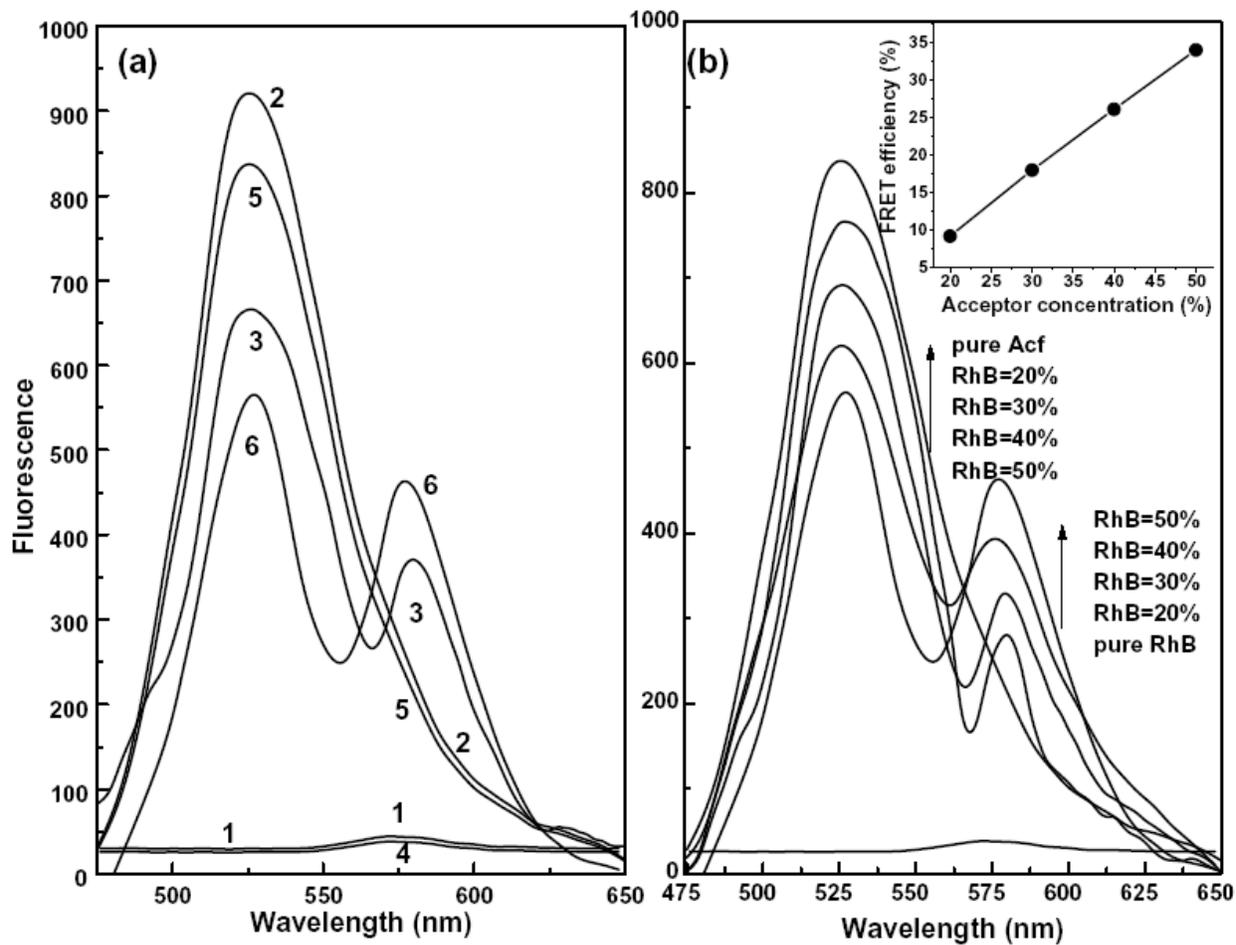

**Fig. 3a and 3b**

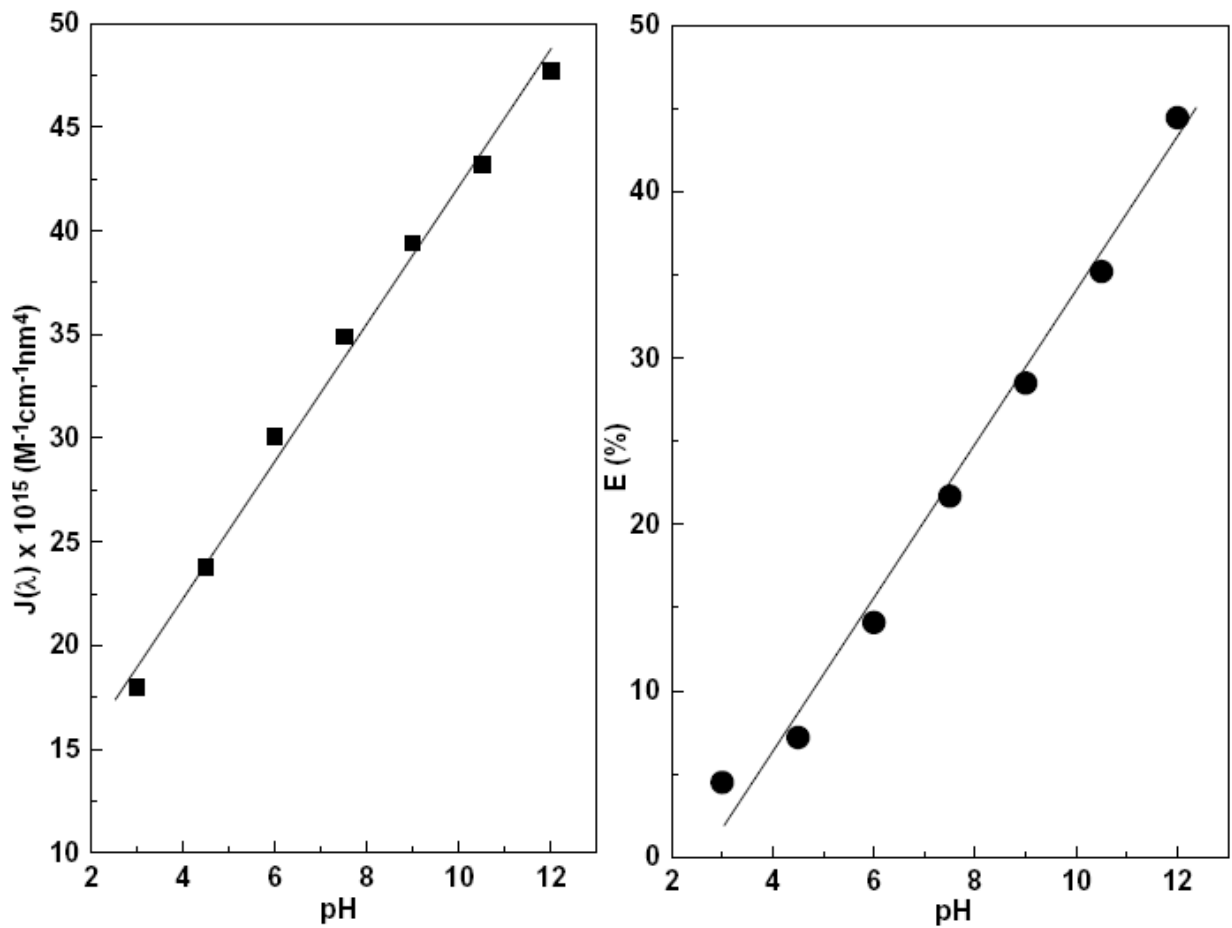

**Fig. 4**

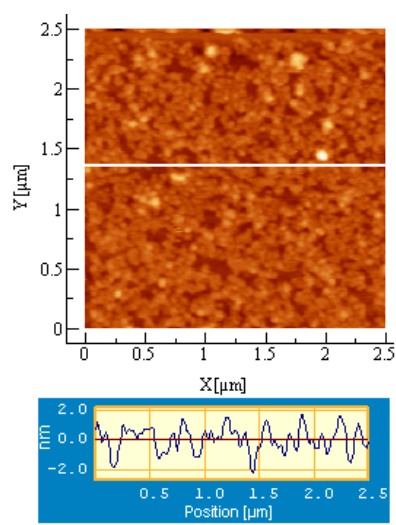

**Fig. 5**